\documentstyle[preprint,aps]{revtex}
\begin{document}
\draft
\begin{title}
{Numerical Computation of Finite Size Scaling Functions:
An Alternative Approach to Finite Size Scaling }
\end{title}
\author{Jae-Kwon Kim, Adauto J.F. de Souza\cite{addr}, and D. P. Landau}
\address
{Center for Simulational Physics, \\
The University of Georgia, Athens, GA 30602}
\maketitle

\begin{abstract}
Using single cluster flip Monte Carlo simulations we accurately determine
new finite size scaling functions which are expressed only in terms the
variable $x = \xi_L / L$, where $\xi_L$ is the correlation length in a finite
system of size $L$. Data for the d=2 and d=3 Ising models, taken at
different temperatures and for different size lattices, show excellent data
collapse over the entire range of scaling variable for susceptibility
and correlation length. From these finite size scaling functions we can
estimate critical temperatures and exponents with rather high accuracy
even though data are not obtained extremely close to the critical point.
The bulk values of the renormalized four-point coupling constant
are accurately measured and show strong evidence for hyperscaling.
\end{abstract}
\section{INTRODUCTION}
True critical phenomena can possibly take place only in the limit
in which the size of the system becomes infinite 
(i.e. the thermodynamic limit).
The singular behavior of a system near a critical point is
characterized by the bulk values of various physical quantities;
it is technically impossible, however, to directly obtain information about
infinite lattices from Monte Carlo simulation.  
Practically, however, it is
not necessary to make the size of the lattice infinite in order to estimate
a thermodynamic value through a Monte Carlo simulation of finite lattices:
The concept of the finite-size-scaling (FSS)\cite{FIS},
introduced to extrapolate the information available from the finite
system to the infinite volume limit, has been remarkably successful.
The most frequent application of FSS\cite{PRIV} has been primarily concerned
with extracting some universal quantities such as the critical exponent $\nu$
or some ratios of the critical exponents, without the knowledge of the bulk
values in the scaling regime.  The standard finite size scaling variable
is $x=tL^{1/\nu}$ with the reduced temperature
$t\equiv |K_{c}-K|/K_{c}$, where $K$ is the inverse coupling $K = J/kT$
and $K_c$ is the inverse critical coupling.  
Of course, use of this variable
presupposes knowledge of the correct critical coupling and the uncertainty
in $K_c$ introduces a source of error into the analysis.  
Another formulation
proposed by Fisher used $\tilde t = |K_{c}^L -K|/K_{c}^L)$, but since this
 requires knowledge
of multiple finite lattice ``critical couplings" it has thus seldom been used.
Some non-universal critical parameters like $K_c$ can also be
calculated by different FSS techniques, e.g. the $4^{th}$ order
 cumulant ratio
method\cite{BIN} or the microcanonical Monte Carlo method\cite{SOU}.

Nonetheless, determining bulk values (i.e. in the thermodynamic limit)
is an important task of MC simulations,
because physical quantities can then be directly compared with experiment.
Also, the variation of a suitable thermodynamic quantity with temperature
near criticality characterizes its critical behavior, even if it is not
describable by a power law.
A criterion telling whether a quantity measured on a finite lattice at a
temperature $T$ is distinguishable from the thermodynamic value (value
in the thermodynamic limit) is the ratio
of the linear size of the lattice ($L$) to the correlation length ($\xi(T)$):
provided $L/\xi(T)$ is sufficiently large, the measured quantity becomes
essentially independent of $L$.  Thus, one needs very large $L$ at temperatures
where $\xi(T)$ becomes large. Unfortunately, in this situation critical
slowing down limits the quality of the data.
Recently, new techniques of FSS have been
introduced\cite{KIME}\cite{SOK} which enable one to extract correct
thermodynamic values indirectly for a variety of physical quantities.
The feature characteristic of these techniques 
is the calculation of some
FSS functions defined in terms of 
a non-conventional FSS variable.

In this paper, we numerically calculate certain 
FSS functions which are
different from the ``usual" ones and extract estimates 
for the values of critical parameters for the two and 
three dimensional Ising models.  
In the next section we provide theoretical background, 
and in the following section
we calculate bulk values of the correlation length ($\xi$), magnetic
susceptibility ($\chi$), and the renormalized four-point coupling
constant ($g^{(4)}$).  We summarize and conclude in the final section.

\section{THEORY AND SIMULATION}
The fundamental assumption of FSS theory\cite{FIS} is that
$A_{L}(t)$, the value of some thermodynamic quantity $A$ on a 
finite lattice of linear size $L$, can be expressed as
\begin{equation}
A_{L}(t)=L^{\rho/\nu} f_{A} (s(L,t)),~~
s(L,t)\equiv L/\xi(t) \label{eq:fun1}
\end{equation}
for a bulk quantity $A$ which has a power-law critical singularity
$A(t) \sim t^{-\rho}$ where $t=(K_c-K)/K_c$.
Eq.(\ref{eq:fun1}) is valid for values
of $L$ and $\xi(t)$ which are large; 
otherwise, there should be corrections to
FSS, which unless explicitly stated are 
ignored throughout this work.

Notice that using the critical form for $\xi$, 
$\xi(t) \sim t^{-\nu}$, we can
rewrite the scaling variable $s(L,t)$ as
\begin{equation}
s(L,t)=(A(t)/L^{\rho/\nu})^{-\nu/\rho},
\end{equation}
so that Eq.(\ref{eq:fun1}) may be rewritten as
\begin{equation}
A_{L}(t)=A(t) {\cal F}_{A}(s(L,t)),    \label{eq:fun2}
\end{equation}
where the relation between the scaling functions $f_{A}$  and 
${\cal F}_{A}$ is given by
\begin{equation}
{\cal F}_{A}(s)=s^{\rho/\nu} f_{A}(s). \label{eq:rel1}
\end{equation}
For $A=\xi$, Eq.(\ref{eq:fun2}) shows that $\xi_{L}(t)/L$ is
just a function of $\xi(t)/L$ and vice versa, and this leads to
the relation
\begin{equation}
A_{L}(t)=A(t){\cal{Q}}_{A}(x(L,t)), \label{eq:fun}
\end{equation}
where $x(L,t)\equiv \xi_{L}(t)/L$ is the ratio of the correlation
length {\bf on a finite lattice} to the lattice size, and
${\cal{Q}}_{A}(x)$ is given by
\begin{equation}
{\cal{Q}}_{A}(x)={\cal F}_{A}(f^{-1}_{\xi}(x)).    \label{eq:rel2}
\end{equation}
Using the same observation, it is trivial to obtain\cite{SOK1}
another equivalent form,
\begin{equation}
A_{bL}(t)=A_{L}(t){\cal{G}}_{A}(x(L,t)),
\end{equation}
where $b$ is a scaling factor and ${\cal{G}}_{A}(x)$ is another
scaling function.

It is evident that given $f_{A}$ one can determine the other
two scaling functions from Eqs.(\ref{eq:rel1}) and (\ref{eq:rel2}),
and all the scaling functions, $f_{A}$, ${\cal{F}}_{A}$,
${\cal{Q}}_{A}$, and ${\cal{G}}_{A}$ should be universal.
It has also been argued\cite{HIL} that a certain asymptotic form
of $f_{A}(s)$ can be expressed in terms of the critical exponent
 $\delta$; by
fitting this functional form one can extract an estimate for 
the critical exponent.

It is worth stressing that use of the scaling function
$\cal{Q}$ rather than $\cal{F}$ would be more convenient in many cases,
particularly because one does not need the bulk correlation length
to define the former.  Note that there is no explicit $t$ dependence of
the scaling variables, so that knowledge of the critical coupling is not
required, and that $x$ becomes independent of $L$ at criticality.
This $L$ independent value of $x$ at criticality, $x_{c}$, which
characterizes a universality class for a given geometry\cite{COM},
forms the upper bound of $x$. 
In other words, the scaling function
$\cal{Q}$ is defined only over $0 \le x \le x_{c}$.  
A priori, the two
limits of the scaling function $\cal{Q}$  are known for 
a continuous phase transition:  
$\lim_{x \to 0} {\cal Q}(x) \to 1$ and
$\lim_{x \to x_{c}} {\cal Q} \to 0$, because $A_{L}$ converges
to its bulk value in the former case while $A(t)$ diverges 
in the latter case with $A_{L}(t)$ finite.  
In general, as we will show in this work,
for $A=\xi$, $\chi$, or $g^{(4)}$, ${\cal{Q}}_{A}(x)$
turns out to be a monotonically decreasing function of $x$.

It is important to realize that the knowledge 
of the scaling function
${\cal Q}$ near $x \simeq 0$ plays as relevant 
role as that near
$x \simeq x_{c}$ to the extraction of necessary information 
of the critical behavior in (deep) scaling region. 
It can be easily seen by noting that
$x(L,t)$ for a fixed temperature arbitrarily close to 
criticality can be made arbitrarily close to zero by 
simply choosing a value of $L$
sufficiently large.

Eqs.(\ref{eq:fun2}) and (\ref{eq:fun}) do not include 
any critical exponents,
so that one might conjecture that their validity can be
extended to non-power-law singularities. 
Although a general proof of this
conjecture is missing, L\"{u}scher\cite{LUS} obtained an
explicit expression for the inverse correlation length (mass gap),
that is consistent with Eq.(\ref{eq:fun}), 
for the two-dimensional (2D)
$O(N)$ ($N >2$) spin models which exhibits an exponential 
critical singularity.
Also, very extensive numerical verification\cite{KIME} of
Eq.(\ref{eq:fun}) was given for $\chi$ and $\xi$ for the 2D $O(N)$
models with N=2 and 3.

In order to define a correlation length, we consider
the Fourier transform of the (connected) two-point correlation function,
\begin{equation}
G({\bf k})\equiv \sum_{{\bf x}}\exp(i k\cdot x)
<S_{0}\cdot S_{{\bf x}}>_{c},
\end{equation}
where $S_{{\bf x}}$ denotes the spin variable at site ${\bf x}$.
When ${\bf x}$ is sufficiently large,
$<S_{0}\cdot S_{{\bf x}}>_{c} \sim e^{-|{\bf x}|/\xi_{L}}$ holds\cite{TGR}, 
so we will have
\begin{equation}
G({\bf k})^{-1}~=~G({\bf 0})^{-1} [ 1+ k^{2}\xi_{L}^{2}+ {\cal O}(k^{4})]
\end{equation}
By choosing ${\bf k}=(2\pi /L,0)$, we obtain
\begin{equation}
\xi_{L}={1 \over 2\sin(\pi/L)} \sqrt{G({\bf 0})/G({\bf k})-1},   \label{eq:cor}
\end{equation}
for values of $L$ that are large enough that terms $\sim k^4$
can be safely ignored.

The renormalized coupling constant $g^{(4)}_{L}$ 
may be defined as\cite{PAT}
\begin{equation}
g^{(4)}_{L}= 3(L/\xi_{L})^{D}~U_{L},
\end{equation}
where $D$ is the lattice dimensionality and the 4th order cumulant
is given by $U_{L}\equiv 1-<S^{4}>/3<S^2>^{2}$,
with $S$ being the order parameter.
The bulk $g^{(4)}$ has a well defined scaling 
behavior\cite{SOK1},
\begin{equation}
g^{(4)}(t) \sim t^{D\nu-2\Delta+\gamma}.
\end{equation}
$g^{(4)}$ describes the non-Gaussian character of the the model,
i.e., only for a Gaussian model does $g^{(4)}(t)$ 
vanish as $t \rightarrow 0$ in the absence of certain 
multiplicative correction to scaling,
implying the violation of the hyperscaling 
relation $D\nu-2\Delta+\gamma >0$.
For a system where the hyperscaling relation is satisfied
(without certain multiplicative correction to scaling as in the 
four dimensional Ising model), 
its bulk value in the scaling regime remains a constant that 
characterizes its universality class.

Employing the single cluster Monte Carlo algorithm\cite{WOL} we
simulated the 2D and
3D Ising models, on the square and simple cubic lattices 
respectively, with
fully periodic boundary condition imposed.  
For each lattice at a given
temperature, we generated up to 30 bins of data 
each of which is composed of
10000 measurements. 
In order to reduce the correlation between data points,
only configurations 3-7 Monte Carlo step apart were considered.
Our quoted errors, which are purely statistical in nature, 
are the standard deviation of the binned values.
Aware of the bad performance of some random number generators
in the context of the single cluster algorithm\cite{DPL}, we have
double-checked our results by
comparing data generated by two different implementations 
of the algorithm
each one using a different kind of random number generator.
Most of data were obtained with a linear congruential random number
generator of the form $x_{i+1}=69069 \times x_{i} +1 ~mod~ (2^{31})$.
The other random number generator was a multiplicative, lagged Fibonacci 
generator of the form
$x_{i+1}=x_{i-4423}\times x_{i-1393},$
which showed a good performance in a single cluster simulational test of
Ising system\cite{COD}.
We observed complete agreement between the two sets of data within our
statistical errors. We therefore, believe that to within error bars quoted here
our data are not biased due to any correlation among the random numbers.
(We also tried the well known R250 routine but the data exhibited some
systematic deviation and we did not consider them in our analysis.) 

\section{RESULTS}
\subsection{2D Ising Model}
We now investigate the finite size behavior for 
a variety of multiplicatively
renormalizable physical quantities\cite{LUS} defined 
on a finite lattice of linear size
$L$, in particular the susceptibility $\chi$ and correlation length $\xi$.
First, we choose a certain $K$ and perform 
measurements of $A_{L}(K)$ for
various values of $L$. 
In Table(1) we present our data for $K=0.425$, with $L$
varying from 16 to 150. 
The reason for starting from  $L=16$ is that near this
$L$ the systematic error in our definition of $\xi_{L}$, Eq.(\ref{eq:cor})
is about $\sim 10^{-2}$ which is comparable 
to our typical statistical errors.
From these data in Table(1) one sees that $\xi$ is 
indistinguishable from bulk value for
$L \ge 80$, to within very small statistical error.  
In terms of the scaling
variables, this means that $s \ge 5.076$ or $x \le 0.196$.
As we stressed earlier, this condition holds 
(in terms of the scaling variables)
regardless of the temperature as long as it remains in the scaling
regime.  (This is indeed the fundamental
statement of FSS.)  
We note that for $\chi$ the thermodynamic condition holds
for a slightly larger value of $s$ (smaller $x$) than that for $\xi$, 
namely $s \ge 6.346$.  
Figure(1) shows the data-collapse for ${\cal F}_{A}(s)$.

From the data in Table(1), we can easily determine ${\cal{F}}_{A}(s)$
and ${\cal{Q}}_{A}(x)$.  In order to satisfy the asymptotic conditions,
for the former one may try either a polynomial function of $1/s$ or $e^{-s}$,
while for the latter a polynomial 
of $x$ or $e^{-1/x}$ may be tried. That is,
\begin{eqnarray}
{\cal{F}}_{A}(s) &=&1+b_{1}/s+b_{2}/s^{2}+\ldots,   \\
{\cal{Q}}_{A}(x)&=&1+b_{1}x+b_{2}x^{2}+\ldots,
\end{eqnarray}
or
\begin{eqnarray}
{\cal{F}}_{A}(s)&=&1+c_{1}e^{-s}+c_{2}e^{-2s}+\ldots,  \\
{\cal{Q}}_{A}(x)&=&1+c_{1}e^{-1/x}+c_{2}e^{-2/x}+\ldots.   \label{eq:form}
\end{eqnarray}
In general, it turns out that for 
the same number of fitting parameters
the polynomial of the exponentials fits better than that of the simple
scaling variables.  
This is especially true for the magnetic
susceptibility and the four-point renormalized coupling constant.  
For instance, by considering terms up to 
the fourth order of the polynomial,
$\chi^{2}/N_{DF}({\rm degree~~of~~freedom})=$ 3.3 and 0.3  respectively
for the ${\cal{Q}}_{\chi}(x)$  assuming 
the simple polynomial and
that of the exponential.  
Considering up to the $e^{-4/x}$ term,  we obtain
$c_{1}=-2.402, c_{2}=-16.338, c_{3}=80.688, c_{4}=-134.6$ for
${\cal{Q}}_{\chi}$ with $\chi^{2}/N_{DF}$=0.31, while they are $-0.768,
-8.490, 31.032$, and $-89.203$ respectively
for ${\cal{Q}}_{\xi}$ with $\chi^{2}/N_{DF}$=0.20.

A priori, the estimates are accurate only for
$x \le \xi_{L_{0}}/L_{0}\equiv x_{0}$, 
with $L_{0}$ denoting the smallest value of
$L$ for $K=0.425$; for the estimate of the coefficients for $x > x_{0}$,
one needs similar data for $A_{L}$ at a larger $K$, which might
modify the values of the coefficients.
Nevertheless, with the information of the finite 
size-scaling function ${\cal{Q}}_{A}$, 
it is now possible to extract accurate bulk data
from the Monte Carlo data on a modestly {\it small} lattice 
provided a data point
(at another temperature) satisfies $x < x_{0}$.  
Our results for $K=0.430,
0.434, 0.436$, and 0.438 are summarized in Table(2).

We notice that the bulk values thus extracted for 
a given $K$ do not change
with respect to $L$, indeed verifying this form of FSS 
for the model (see Figure(2) also).
We also note that the values of the bulk $\xi$ thus extracted 
are in excellent agreement
with the corresponding exact values given by the formula
\begin{equation}
1 /\xi(K) =\ln(\coth K)-2K,  \label{eq:excor}
\end{equation}
within typically less than 0.5\% of the statistical errors.  
As a test of our $\chi$, we fitted the data over the range 
from $K=0.425$ to 0.438  to
\begin{equation}
\chi \sim t^{-\gamma}.
\end{equation}
The best $\chi^{2}$ fit gives $K_{c}=0.44070(5)$ and
$\gamma=1.755(9)$ ($\gamma$=1.752 by fixing $K_{c}$ to the
exact critical point), being extremely close to the exact values.
Since all of our data used for the analysis were for $t > 0.006$,
the quantity of the result is surprisingly good.

\subsection{The 3D Ising Model}
We begin with our Monte Carlo measurement at $K=0.220$, 
the results of which are summarized in Table(3).  
We observe that the thermodynamic condition for
the $\xi$ is almost satisfied for 
$L/\xi_{L} \approx 60/10.89 \approx 5.5$.
Nevertheless,
we also note that $\chi$ and $g^{(4)}$ increase, albeit very slowly,
beyond this value. This is another indication that certain quantities
converge to the thermodynamic value more 
slowly than the correlation length.

Assuming that $A_{L}(K=0.220)$ reaches its bulk value for $L=70$, 
we obtain
\begin{equation}
c_{1}\simeq -0.418,\hfill c_{2}\simeq -18.83, \hfill c_{3}\simeq 99.38,
\hfill c_{4}\simeq -436.4,
\end{equation}
\begin{equation}
c_{1}\simeq -0.607,\hfill  c_{2}\simeq -56.31, \hfill c_{3}\simeq 416.75, 
c_{4}\simeq -1399.9
\end{equation}
\begin{equation}
c_{1}\simeq -7.238,\hfill c_{2}\simeq 21.42, \hfill  c_{3}\simeq -16.09,
\hfill c_{4}\simeq 3.67,
\end{equation}
respectively for ${\cal{Q}}_{\xi}(x)$, ${\cal{Q}}_{\chi}(x)$, and
${\cal{Q}}_{g^{(4)}}(x)$, over the range $0 \le x \le x_{0}\equiv 0.491$.

Based on the knowledge of ${\cal Q}_{A}(x)$, 
we calculate the bulk values of
the correlation length, magnetic susceptibility, 
and four-point renormalized
coupling constant for various $K$ up to $K=0.2212$. 
The largest value of $L$
we simulated for the calculation is just 64. 
Obviously, the computation of
the bulk value at a larger $K$ (than 0.2212) requires 
a larger value of $L$ in
order to keep the value of $x$ smaller than $x_{0}$.  
One way to avoid the
need for a larger $L$ is to repeat the measurement 
of $A_{L}$ at a slightly
larger value of $K$, e.g. $K=0.2212$; this will extend the range of $x$
over which ${\cal Q}_{A}$ is accurately computed. 
In the region where
${\cal Q}_{A}(x) \approx 0$, however, one needs 
extremely precise measurements
to reduce the errors in the estimates of the bulk values\cite{COM2}.

A summary of our results is shown in Table(4).  We see that
the four- point renormalized coupling constant
remains unchanged, i.e., $g^{(4)} \approx 24.5$, for $K\ge 0.2206$.
Its slow variation for $K < 0.2206$ may be due to the presence of
correction to scaling.
Hence the hyperscaling relation is indeed satisfied, 
confirming the previous verification (within rather larger statistical 
errors though) based on the traditional 
Monte-Carlo measurement\cite{PAT}\cite{TSY}.
We would like to stress, however, 
that it was not possible to measure
$g^{(4)}$ beyond $K=0.2206$ in Ref.\cite{PAT}, 
even using $L$ as large as $L=90$.  
The bulk $\chi$ and $\xi$ thus extracted are compared with those
traditionally obtained, again yielding remarkable agreement (see Table(4)).
Figure(3) exhibits excellent data-collapse for the finite size
scaling function ${\cal Q}_{A}(x)$ for $A=
\chi, ~~\xi$, and $g^{(4)}$.

In order to determine $K_{c}$, $\nu$, and $\gamma$
we fitted our bulk data over the range $0.217 \le K \le 0.2212$
to the simple power-law singularity.
We fixed the critical point in the fit, and then repeated 
the fit for several
different fixed critical points.  The results, shown in Fig. 4,
indicate that the $\chi^{2}$ values of the $\xi$ and $\chi$
data favor the range of $K_{c}$
over $ 0.221640 \le K_{c} \le 0.221670$, 
being consistent with recent other
results\cite{LAN}.  
The empirical formulae we obtained from the best fit are as follows:
\begin{eqnarray}
\xi &=& C_{\xi} (|K-K_{c}|/K_{c})^{-\nu},~~ C_{\xi}=0.4710,
           ~~ K_{c}=0.221658,~~\nu=0.6418 \\
\chi &=& C_{\chi} (|K-K_{c}|/K_{c})^{-\gamma},~~C_{\chi}=1.0892,
          ~~K_{c}=0.221646, ~~\gamma=1.2388 
\end{eqnarray}
The value of $\nu$ is larger by approximately 1.5 percent 
than those extracted by most other methods, 
while the value of $\gamma$ agrees up to $\sim 10^{-3}$.
The effect of including the term of the confluent correction 
to the scaling turns out to be minimal: the confluent 
correction-term would usually be important when data with rather 
smaller bulk values of the correlation length are considered. 
Given the modest size lattices
used and the distance from the critical points at which 
the actual measurements were made ($t > 0.002$), we find that the 
agreement with high resolution studies to be
extremely gratifying.

\section{DISCUSSION AND CONCLUSION}
In this paper we computed an alternative  
finite size scaling function, defined
in terms of the scaling variable $\xi_{L}/L$, 
for the 2D and 3D Ising models.
This type of finite size scaling function has the 
advantage of being defined even for small values of $L$,
without any prior information on the critical behavior
of the system.  
Thus, our procedure can test FSS itself
by means of data collapse 
as far as $\xi_{L}$ can be 
accurately measured; in this manner, we observed that
the effect of the violation of FSS is 
negligibly small at least for $L \ge 16$ for the two and three 
dimensional Ising models.
We illustrated how the function can be used
for the extraction of correct bulk values near criticality, 
and that it
can be used in extracting accurate critical parameters
provided the values of the correlation length 
are sufficiently large, i.e. approximately $\xi \ge 5$. 

One might wonder if this technique requires an 
accurate bulk value of a
physical quantity $A$ at least at a point of temperature.  
As far as the extraction of
critical parameters is concerned, however, 
this is not necessarily the case.
To see this, imagine that we start with a fake 
`bulk value' $A^{\prime}(t)$ instead of correct one $A(t)$. 
The scaling function ${\cal Q^{\prime}}$ defined
in terms of the fake bulk value,  
${\cal Q^{\prime}}_{A} \equiv A_{L}(t)/A^{\prime}(t)=
{A(t) \over A^{\prime}(t)} {\cal  Q}_{A}$, 
simply rescales the correct scaling
function by a constant; 
accordingly, every bulk value calculated at any
other temperature
using ${\cal Q}^{\prime}_{A}$  rescales the correct 
one with an overall
factor ${A^{\prime}(t) \over A(t)}$. 
This overall factor is unimportant
for the extraction of the critical behavior.  
One can thus repeat our analysis arbitrarily close 
to a critical point, where the effect of correction to scaling 
can be arbitrarily small. We anticipate that such an analysis 
will yield extremely accurate estimate of the critical parameters,
and our study on the 3D Ising model along this line is under way.
(This observation is also important for some calculations 
of lattice gauge theory. For example, in full lattice QCD, 
the computation of the ratio of the mass of
various lattice `hadrons' is of primary concern, 
and for this purpose the overall factor is simply unimportant.)

We would like to stress again that the technique we 
have illustrated is extremely general; 
it holds regardless of the functional form of the critical
singularity, and irrespective of the quantity as far as it is
multiplicatively renormalizable.  
We demonstrated this point taking the
example of the four-point renormalized coupling constant, 
whose bulk value is notoriously difficult to measure 
through traditional Monte Carlo simulation\cite{TSY}\cite{BAK}.
Although  our current estimate of the critical parameters cannot 
compete with the highest resolution Monte Carlo studies, our
estimates are really 
surprisingly good considering how far from the critical 
point the data are taken.  Thus,
although the determination of a finite lattice 
correlation function is needed,
this method offers a simple alternative to
standard finite size scaling methods. \\

\acknowledgments 
{This research was supported by the CNPq and by NSF grant DMR-9405018.
  We are grateful to K. Binder, B. D\"{u}nweg, and A. Ferrenberg for helpful
  comments.}

\begin{table}
\caption {$A_{L}$ as a function of $L$ at $K$=0.425.
$\xi_{\infty}(K$=0.425)=15.7582... from Eq.(17).  
Note that our $\xi_{L}$ converges to its bulk value for 
$L \ge 80$, within the small statistical errors, and that
the value of $x$ monotonically decreases with $L$.}
\begin{tabular}{cccc}
$L$	&$\xi_{L}$  &$x$  &$\chi_{L}$ \\
\hline
16.   & 9.83(3)    &0.614(2)	&102.4(2) \\
18.   &10.56(4)    &0.587(2)	&119.8(3) \\
20.   &11.18(5)    &0.559(2)	&137.0(6)\\
22.   &11.74(4)    &0.534(2)	&153.5(5) \\
25.   &12.44(5)    &0.498(2)    &177.1(7) \\
27.   &12.85(5)    &0.476(2)    &191.9(7) \\
30.   &13.37(6)    &0.446(2)    &212.7(6) \\
32.   &13.69(7)    &0.428(2)    &224.7(9) \\
34.   &13.92(6)    &0.409(2)    &235.2(9) \\
36.   &14.19(6)    &0.394(2)    &246.6(8) \\
40.   &14.54(6)    &0.363(2)    &264.9(6)\\
50.   &15.19(8)    &0.304(2)    &296.0(1.0) \\
60.   &15.40(6)    &0.257(1)    &312.5(1.0) \\
70.   &15.62(7)    &0.223(1)    &321.3(1.4) \\
80.   &15.71(8)    &0.196(1)    &326.6(1.3)\\
100.  &15.75(10)   &0.157(1)    &329.4(1.8) \\
110.  &15.77(10)   &0.143(1)    &331.0(1.5) \\
150. &15.75(14)    &0.105(1)    &331.4(1.6) \\
\end{tabular}
\end{table}

\begin{table}
\caption{The extraction of the $\xi$ and $\chi$ for the 2D Ising model
         based on the computation of the ${\cal{Q}}_{A}(x)$. 
         The `{\bf ave.}' for each $K$ denotes the average over 
         the extracted bulk values from
         the different values of $L$($x$).}
\begin{tabular}{cccccccc}
$K$   &$L$  &$\xi_{L}$  &$x$  &$\chi_{L}$   &$\xi$   &$\chi$   \\
\hline
0.430 & 20. &13.00(7) &.650  &158.5(4)   &23.16(17)   &650.4(16.2) \\
      & 30. &16.67(7) &.556  &274.5(4)   &23.33(11)   &653.2(5.4) \\
      & 40. &18.97(8) &.474  &377.3(6)   &23.20(9)   &647.5(4.2) \\
      & 50. &20.57(8) &.411  &461.1(8)   &23.31(8)    &650.2(3.8) \\
      & 60. &21.57(8) &.359  &524.5(1.5) &23.30(7)   &651.0(5.5) \\
      & 80. &22.57(10) &.282 &597.0(2.2) &23.23(8)  &649.7(6.3) \\
      &100. &22.89(11) &.229 &625.3(2.3) &23.15(9)  &646.6(5.9) \\
      &120. &23.12(17) &.193 &638.2(5.0) &23.23(13) &647.2(12.4) \\
{\bf ave.}  &-    &-     & -    & - &{\bf 23.21(11)} &{\bf 649.5(2.2)} \\
{\bf exact} &-    &-     &-     &-  & \bbox{23.22}     &- \\ \\ 

0.434 & 80. &32.94(7) &.412  &1047.5(2.5) &37.34(7) &1478.6(11.9) \\
      &160. &36.77(9) &.230  &1426.5(9.0) &37.19(14) &1476.0(23.1) \\
{\bf ave.}  &-    &-    &-     &-  &{\bf 37.27(11)} &{\bf 1477.3(1.8)} \\
{\bf exact} &-    &-    &-     &-  &\bbox{37.21}     &-  \\ \\

0.436 &50. &31.58(9) &.632  &761.7(1.4) &53.10(20) &2745.0(43.7) \\
      &60. &35.35(11) &.589 &977.4(2.1) &53.18(19) &2743.8(39.8) \\
      &70. &38.54(12) &.551 &1186(3)    &53.41(18) &2759.0(37.7) \\
      &80. &41.23(13) &.515 &1389(3)    &53.65(17) &2779.4(31.8) \\
      &90. &43.08(7) &.479  &1563(2)    &53.01(8)  &2724.8(17.5) \\
      &100.&45.04(13) &.450 &1746(2)    &53.37(14) &2751.1(24.7) \\
      &120.&47.96(14) &.400 &2033(5)    &53.69(13) &2776.8(22.4) \\
      &160.&50.55(20) &.316 &2384(7)    &52.96(17) &2724.1(22.9) \\
{\bf ave.}  &-   &-     &-    &-    &{\bf 53.30(28)} &{\bf 2750.6(20.8)} \\
{\bf exact} &-   &-     &-    &-        & \bbox{53.16}     &- \\ \\

0.438 &80. &51.88(13) &.649 &1777(4)  &91.97(31) &7209.7(142.4) \\
      &160.&76.31(34) &.477 &4259(13) &93.66(39) &7378.2(93.0) \\
{\bf ave.}  &-   &-     &-     &- &{\bf 92.82(1.20)}&{\bf 7294.0(119.1)} \\
{\bf exact} &-   &-     &-     &- & \bbox{92.86}      &-  \\ \\
\end{tabular}
\end{table}

\begin{table}
\caption{$A_{L}(K)$ at $K=0.220$ for the 3D Ising model.}
\begin{tabular}{ccccc}
$L$     &$\xi_{L}$ &$x$  &$\chi_{L}$   &$g^{(4)}_{L}$ \\
\hline
16.   &7.85(2)   &0.491(1) &228.8(7) &9.52(5) \\
20.   &8.85(2)   &0.443(1) &298.1(7) &11.43(5) \\
24.   &9.56(2)   &0.398(1) &351.8(1.1)&13.44(7) \\
30.   &10.20(3)  &0.340(1) &407.1(1.1)&16.5(1) \\
36.   &10.56(2)  &0.293(1) &439.5(1.2)&19.2(2) \\
40.   &10.68(3)  &0.267(1) &455.2(1.5)&21.4(3) \\
50.   &10.85(3)  &0.217(1) &467.9(1.3)&23.5(5) \\
60.   &10.89(3)  &0.182(1) &472.3(0.8)&24.2(5) \\
70.   &10.91(3)  &0.156(1) &473.0(1.1)&24.7(5) \\	
\end{tabular}
\end{table}

\begin{table}
\caption{The extraction of the $\xi$ and $\chi$ for the 3D Ising model.
         The uncertainties in the values of $x$ are not considered 
         for our error estimates.
         Traditional Monte Carlo measurements are in rows labeled MC.}
\begin{tabular}{ccccccccc}
$K$    &$L$   &$\xi_{L}$   &$x$   &$\chi_{L}$ &$g^{(4)}_{L}$ &$\xi$   &$\chi$
&$g^{(4)}$ \\
\hline
0.217  &16. &5.30(1)  &.331  &114.6(2)   &17.7(1) &5.62(1)    &130.8(2)    &25.5(1) \\
       & 20.&5.49(1)  &.275  &124.5(4)   &21.2(2) &5.63(1)    &130.7(4)    &25.4(1) \\
       & 28.&5.60(1)  &.200  &130.1(1)   &24.6(2) &5.62(1)    &131.0(3)    &25.8(2) \\
{\bf ave.}   &-   &-        &-     &-          &-       &\bbox{5.62(1)}    &\bbox{130.9(4)}    
       &\bbox{25.6(3)} \\ \\

0.219  & 20. &7.33(2) &.366  &212.4(7)   &15.3(2) &8.04(2)    &262.2(9)    &24.9(6) \\
       & 30. &7.88(3) &.263  &251.5(1.5) &21.8(5) &8.03(4)    &260.9(1.6)  &25.7(6) \\
{\bf ave.}   & -   & -      &-     &-          &-       &\bbox{8.03(4)}    &\bbox{261.6(2.1)}  &\bbox{25.3(1.1) } \\ 
MC     & 50  &-       &-     &-          &-       &8.1        &263         &26 \\ \\

0.2203 &30. &11.16(8)  &.372  &481.4(5.3) &14.8(1) &12.31(9)  &602.7(6.6)  &24.6(2) \\
       &40. &12.00(3)  &.300  &561.4(2.6) &19.3(2) &12.44(3)  &608.2(3.1)  &25.1(4) \\
{\bf ave.}   &-   &-        &-     &-          &-       &\bbox{12.38(15)}  &\bbox{605.5(8.8)}  &\bbox{24.9(7)} \\ \\

0.2206 &24. &11.20(5) &.467  &467.4(2.8) &10.3(2) &14.55(6)   &834.2(5.0)  &24.5(2) \\
       &30. &12.49(10)&.416  &589.2(2.7) &12.5(2) &14.66(12)  &844.4(3.9)  &24.6(2) \\
       &36. &13.28(5) &.369  &673.8(3.6) &14.9(3) &14.60(5)   &836.8(4.5)  &24.5(3) \\
       &40. &13.70(4) &.343  &718.0(3.4) &16.3(2) &14.66(4)   &838.4(5.0)  &24.4(3) \\
       &50. &14.28(5) &.286  &789.7(4.0) &19.4(3) &14.68(5)   &839.7(5.3)  &24.2(4) \\
       &60. &14.38(3) &.240  &809.9(1.6) &22.4(3) &14.53(5)   &827.6(2.6)  &25.1(4) \\
{\bf ave. }  &-   &-        &-     &-          &-       &\bbox{14.61(6)}  &\bbox{837.0(5.7)} 
          &\bbox{24.6(3)} \\
MC     &75  &-        &-     &-          &-       &14.5       &828         &24
\\ \\

0.2210 &40. &16.75(3) &.419  &1045.0(3.1) &12.3(1)&19.73(5)   &1511.5(5.5) &24.5(2) \\
       &50. &18.15(4) &.363  &1245.7(4.1) &15.1(2)&19.82(8)   &1524.3(7.1) &24.4(4) \\
{\bf ave.}   &-   &-        &-     &-           &-      &\bbox{19.77(12)}  &\bbox{1518.(15.)} &\bbox{24.5(3)} \\ \\

0.2212 &56. &21.98(7) &.393  &1796.0(10.5)&13.6(1)&24.89(10)   &2382(14)  &24.5(2) \\
       &64. &22.89(7) &.358  &1971.6(10.4) &15.5(2) &24.86(10) &2382(13)  &24.4(4) \\
{\bf ave.}   &-   &-        &-     &-            &-       &\bbox{24.88(12)} &\bbox{2382(14)}  &\bbox{24.5(3)} \\

\end{tabular}
\end{table}

\vspace{1cm}
{\bf Figure Captions} \\

{\bf Fig.(1)}: ${\cal F}_{A}(s)$ for the 2D Ising model. 
           Each `curve' demonstrates the data collapse
           for two different values of $K$. 
           Note that the lower curve converges
           to ${\cal F}_{A}(s)$=1, i.e., thermodynamic limit
            more slowly than the upper curve. \\
{\bf Fig.(2)}: ${\cal Q}_{A}(x)$ for the 2D Ising model.  \\
{\bf Fig.(3)}: ${\cal Q}_{A}(x)$ for the 3D Ising model. 
               The solid lines are from the best fits. \\
{\bf Fig. (4)}: The resulting estimates for the critical exponents
$\gamma$ and $\nu$ for different choices of the critical coupling $K_c$.
The correspondent errors are given by the light dotted lines. Solid
(dashed) curves correspond to $\chi$ ($\xi$) data. The shadings show
acceptable values for the critical parameters. On the bottom: $\chi^2$
plotted against $K_c$; vertical arrows locate our best estimates for
$K_c$, whereas the horizontal ones indicate the error bars. \\


\begin{thebibliography}{99}
\bibitem{COM} For the 2D square lattice and 3D simple cubic lattice Ising
        models, the values of $x_{c}$ are numerically computed\cite{PAT},
        $x_{c}\approx 0.91(1)$ and 0.64(2) 
        respectively for the former and latter.
\bibitem{COM2} In this manner, $\xi$ was measured 
         up to $\xi \simeq 1.5\times 10^{4}$
         for the 2D $O(3)$ model\cite{KIME}.
\end{thebibliography}
\end{document}